# Equivalence of "Reversible" and "Irreversible" Entropy Modeling

Martti Pekkanen
Department of Chemical Engineering, Helsinki University of Technology, Espoo, Finland
Present address: SciTech-Service Oy, Helsinki, Finland
martti.pekkanen@scitech.fi

## Abstract

There are currently two – main, continuum – models of entropy: a "reversible", Clausius entropy model and an "irreversible", Onsager-Prigogine entropy model.

It is shown that the equations of the "reversible" and the "irreversible" entropy models are equivalent with respect to entropy accumulation – which entails same values of entropy change and, thus, same values of entropy.

The equivalence contradicts the "second fundamental equation" of the "reversible" entropy model, $dS = dQ/T$, holding true for "reversible" phenomena, only.

The equivalence conforms with entropy history independence, which entails that equation $dS = dQ/T$ must hold true for not "reversible" phenomena, also.

Several examples – e.g. by commercial engineering software – show that equation $dS = dQ/T$ holds true for not "reversible" phenomena, also.

The two results of this analysis – the equivalence of the two entropy models and the falsity of "reversibility" – would be falsified by evidence for equation $dS = dQ/T$ not holding true for not "reversible" phenomena. No such evidence has been presented.

The main consequence of the equivalence of the two entropy models is a signification simplification of the language used in the context of entropy modeling.

# 1 INTRODUCTION

## 1.1 Two Models of Entropy

There are currently two – main, continuum – models of entropy: the "reversible" model of Clausius [1-6] and the "irreversible" model of Onsager [7] and Prigogine [8].

The "reversible" entropy model is based on the "second fundamental equation" of the "mechanical theory of heat", Clausius [4], p. 366, [6], p. 110,

$$dS = \frac{dQ}{T} . \tag{II}$$

For the same physical situation, the "irreversible" entropy model, de Groot and Mazur [9], p. 24, [1]

$$\frac{\partial S'''}{\partial t} = -\nabla \cdot \frac{\tilde{\mathbf{Q}}''}{T} - \frac{1}{T^2}\tilde{\mathbf{Q}}'' \cdot \nabla T + \frac{\tilde{Q}'''_{GEN}}{T} . \tag{1}$$

## 1.2 Equivalence

In the context of equation (1), consider the two equations

$$\nabla \cdot \frac{\tilde{\mathbf{Q}}''}{T} = \frac{\nabla \cdot \tilde{\mathbf{Q}}''}{T} - \frac{\tilde{\mathbf{Q}}'' \cdot \nabla T}{T^2} , \tag{2}$$

$$\tilde{Q}'''_{ABS} = -\nabla \cdot \tilde{\mathbf{Q}}'' + \tilde{Q}'''_{GEN} , \tag{3}$$

the former of which holds true for any vector $\mathbf{Q}$ and any scalar $T$, while the latter for heat absorption $Q_{\text{ABS}}$, heat flow $\mathbf{Q}$, and heat generation $Q_{\text{GEN}}$, only.

Now, using equations (2) and (3) in equation (1) gives

$$\frac{\partial S'''}{\partial t} = \frac{\tilde{Q}'''_{ABS}}{T} = \frac{-\nabla \cdot \tilde{\mathbf{Q}}'' + \tilde{Q}'''_{GEN}}{T} = -\nabla \cdot \frac{\tilde{\mathbf{Q}}''}{T} - \frac{1}{T^2}\tilde{\mathbf{Q}}'' \cdot \nabla T + \frac{\tilde{Q}'''_{GEN}}{T} , \tag{4}$$

in which the leftmost equality expresses the *same proposition* as equation (II), which entails that equations (II) and (1) are *equivalent*.

However, according to Clausius [4], p. 366, [6], p. 110, equation (II) holds true for "reversible" phenomena, *only*, while equation (1) is taken to be not thus restricted. Accordingly "reversibility" contradicts the equivalence.

## 1.3 "Reversibility" and Entropy History Independence

According to Clausius [1-6], equation (II) holds true for "reversible" phenomena and for "reversible" phenomena *only*, which two propositions may be expressed as

$$dS_{\text{"rev"}} = \frac{dQ_{\text{"rev"}}}{T} , \tag{5}$$

$$dS_{\text{not "rev"}} \neq \frac{dQ_{\text{not "rev"}}}{T} . \tag{6}$$

---

[1] A tilde denotes a time rate of a quantity and primes denote a quantity per area or volume.

Now, for a "reversible" phenomenon and a not "reversible" phenomenon, with *the same amount of heat*, i.e.

$$dQ_{\text{"rev"}} = dQ_{\text{not "rev"}}\,. \tag{7}$$

equations (5) and (6) entail that

$$dS_{\text{"rev"}} \neq dS_{\text{not "rev"}}\,. \tag{8}$$

This is in contradiction with entropy being "a magnitude which depends only on the present existing condition of the body, and not upon the way by which it reached the latter", Clausius [4], p. 355, [6], p. 90, i.e. with *entropy being history independent* ("path independent", "state function").

Thus, if entropy history independence is true, "reversibility" is false – i.e. it is false that equation (II), $dS = dQ/T$, holds true for "reversible" phenomena, *only*.

### 1.4 Outline

First, the "reversible", Clausius entropy model is considered. The result is that "reversibility" – as a restriction of equation (II), $dS = dQ/T$, to "reversible" phenomena – is false. The falsity follows from entropy history independence. The falsity is corroborated by several examples of equation (II) holding true for not "reversible" phenomena.

Second, the "irreversible", Onsager-Prigogine entropy model is considered. The result is that entropy generation by heat flow and entropy flow by heat flow divided by temperature are the essential concepts of the model.

Finally, it will be shown that the Clausius entropy model and the Onsager-Prigogine entropy model are equivalent with respect to entropy accumulation

### 1.5 Balance

The basic conceptual tool of the analysis is the balance proposition, $ACC = (IN\text{-}OUT) + GEN$, presented in appendix A. The heat balance, equation (3), is essential for the equivalence of the two entropy models.

### 1.6 Justification and Falsification of the Results

The result of this analysis that "reversibility" is false is justified by entropy history independence and is corroborated by ample evidence – e.g. numerical calculations by commercial engineering design software – for equation (II), $dS = dQ/T$, *holding true for not "reversible" phenomena*.

The two results of this analysis – the falsity of "reversibility" and the equivalence of the two entropy models – would be falsified by evidence for equation (II), $dS = dQ/T$, *not holding true for not "reversible" phenomena*. *No such evidence* has been presented – neither by Clausius [1-6] nor by later treatises of entropy modeling.

# 2 CLAUSIUS ENTROPY MODEL

## 2.1 "Reversibility"

The "mechanical theory of heat" of Clausius [1-6] presents the "first fundamental equation" (I)[2] and the "second fundamental equation" (II), Clausius [4], p. 366, [6], p. 110, of

$$dU = dQ + dW \ , \tag{I}$$

$$dQ = TdS \ . \tag{II}$$

Clausius [6], p. 110:

> "Both equations relate to an indefinitely small alteration of condition in the body, and in the latter it is further assumed that this alteration is affected in such a way as to be reversible. For the truth of the first equation this assumption is not necessary […]"

The scope of the "mechanical theory of heat" of Clausius [1-6] are the *heat and work phenomena* of *closed systems*.

According to Clausius [4], p. 133, [6], p. 212, phenomena with *non-zero temperature difference* and with *heat generation* are not "reversible".

Thus, the essence of "reversibility" is the *restriction of the domain* of the "second fundamental equation (II), $dS = dQ/T$, to "reversible" heat and work phenomena within the scope of the "mechanical theory of heat" of all heat and work phenomena.

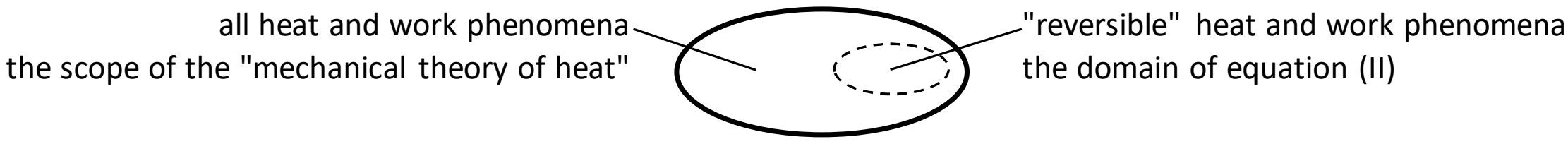

**Figure 1. The Domain of Equation (II) - According to Clausius [1-6]**

## 2.2 Origins of "Reversibility"

The derivations of equations (I) and (II), Clausius [1-6], are based on consideration of *heat engine cycles*, with explicit reference to Carnot [10], Clausius [4], p. 379, [6], p. 23.

The derivation of equation (I), Clausius [1], pp. 379-384, [4], pp. 23-28, refers to operation in forward direction, *only*, while the derivation of equation (II), Clausius [2], pp. 487-501, [3] pp. 386-387, [4], pp. 116-129; 354-355, explicitly refers to operations both in forward and inverse directions. This difference in the derivations *may explain* why equation (I) is not taken and equation (II) is taken to be restricted to "reversible" phenomena by Clausius [1-6].

For the heat engine cycles Carnot [10] makes two assumptions:[3]

1. *Zero temperature difference* for heat transfer in the isothermal steps, Carnot [10], pp. 52-53.
2. *No losses* – "of ether motive power or caloric" – in the heat engines, Carnot [10], p. 54.

These *may explain* why "reversibility" excludes phenomena with *non-zero temperature difference* and with *heat generation* from the domain of equation (II), $dS = dQ/T$, Clausius [4], p. 133, [6], p. 212.

---

[2] A sign changed in (I) and (II) in an "equivalent" form.

[3] Carnot [10] does not *justify* the first assumption, which *may be* based on simplicity of presentation.

## 2.3 Entropy History Independence and "Reversibility"

According to Clausius [3], p. 387, [4], p. 355, [5], p. 94, [6], p. 90, entropy is "a quantity, which depends only on the present condition of the body, and is altogether independent of the way in which it has been brought into that condition", i.e., *entropy is history independent* ("path independent", "state function").

According to Clausius [1-6], equation (II), $dS = dQ/T$, *holds true for "reversible" phenomena*, which may be expressed as

$$dS_{\text{"rev"}} = \frac{dQ_{\text{"rev"}}}{T}, \tag{5}$$

and holds true for "reversible" phenomena, *only* – i.e. *does not hold true for not "reversible" phenomena* – which may be expressed as

$$dS_{\text{not "rev"}} \neq \frac{dQ_{\text{not "rev"}}}{T}. \tag{6}$$

Now, for a "reversible" phenomenon and a not "reversible" phenomenon, with *the same amount of heat*, i.e.

$$dQ_{\text{"rev"}} = dQ_{\text{not "rev"}}. \tag{7}$$

equations (5) and (6) entail that

$$dS_{\text{"rev"}} \neq dS_{\text{not "rev"}}. \tag{8}$$

This is in contradiction with *entropy history independence*.

Thus, if entropy history independence is true, "reversibility" is false.[4]

## 2.4 Justification of "Reversibility"

The essence of "reversibility" is that equation (II), $dS = dQ/T$, holds true for "reversible" phenomena, *only* – i.e. *does not hold true for not "reversible" phenomena*.

Clausius [1-6] do not present *any evidence* for "reversibility, i.e. for equation (II), $dS = dQ/T$, *not holding true for not "reversible" phenomena*.

Accordingly, the truth of "reversibility" rests solely on the derivations by Clausius [1-6] and the assumptions used in the derivations.

For example, the derivations use the assumption of "reservoirs" with temperatures equal to the temperatures of the working fluid during the isothermal steps. However, in the same context, *temperatures of the "reservoirs" are found irrelevant*, Clausius [2], p. 501, [4], p. 130, [5], p. 111, [6], p. 106, see appendix B.

---

[4] Instead of equation (5), it is often written $dS = dQ_{\text{rev}}/T$, which, taking entropy history independence and $dQ_{\text{rev}} = dQ_{\text{not rev}}$, gives $dS = dQ_{\text{not rev}}/T$, which leads to the same conclusion. See appendix C, also.

## 2.5 Falsification of "Reversibility"

The essence of "reversibility" is that equation (II), $dS = dQ/T$, holds true for "reversible" phenomena, *only* – i.e. *does not hold true for not "reversible" phenomena*.

The essence of "reversibility" entails a contradiction with *entropy history independence*, which, if true, falsifies "reversibility" – see equations (5)-(8).

The falsity of "reversibility" is corroborated by the contradiction about the temperatures of the "reservoirs" in Clausius [1-6] – see appendix B.

The falsity of "reversibility" is corroborated by *textbooks accounts* of the calculation of entropy change of not "reversible" phenomena using equation (II) – see appendix C.

The falsity of "reversibility" is corroborated by the ubiquitous use of equation (II) for not "reversible" phenomena in *engineering practice* – see appendix D.

The falsity of "reversibility" is corroborated by *conceptual and numeric examples* – including cycles – of equation (II) holding true for not "reversible" phenomena, also – see appendix E.

Finally, the falsity of "reversibility" is corroborated by the *non-existence of evidence* for equation (II) *not holding true for not "reversible" phenomena.*

The conclusion is that "reversibility" is false.

## 2.6 Consequences of the Falsity

The main consequence of the falsity – and thus irrelevance – of "reversibility" is a signification simplification of the language used in the context of entropy modeling.

The special items of cycle efficiency, inequalities and the principle of the increase of entropy, and equilibrium are considered in appendix F.

# 3 ONSAGER-PRIGOGINE ENTROPY MODEL

## 3.1 “Irreversibility”

Onsager [7], p. 407:[5]

> […] the rate of production of entropy per unit volume of the conductor equals
>
> $$\frac{dS}{dt}=\frac{1}{T}\left(J_1X_1+J_2X_2\right)[...]\text{"}$$

in which $J$ are “flows” and $X$ are “forces” – which concepts are not allowed by the “reversible” entropy model of Clausius [1-6].

Because “reversibility” is false and thus irrelevant, the “reversible” (“equilibrium”) and “irreversible” (“nonequilibrium”) entropy models are called the Clausius and the Onsager-Prigogine entropy models, respectively.

## 3.2 Onsager-Prigogine Entropy Model

The fundamental equation[6] of the Onsager-Prigogine entropy model is, Onsager [7], p. 407, Prigogine, [8], p. 40,[7]

$$\sigma=\frac{1}{T}\left(J_1X_1+J_2X_2\right), \tag{9}$$

in which $J$ are “flows”, $J_1$ the electric current and $J_2$ the heat flow, and $X$ are “forces”.

Next, *first*, the *dimensional homogeneity* of this fundamental equation requires that, Onsager [7], p. 406: [8]

> “[…] the ‘force’ that drives the heat flow is
>
> $$X_2=-\frac{1}{T}\operatorname{grad}T\text{"}$$

Using this in equation (9) gives the entropy generation by heat low

$$\sigma_{J_2}=\frac{1}{T}J_2X_2=-\frac{1}{T^2}J_2\operatorname{grad}T, \tag{10}$$

Accordingly, the dimensional homogeneity of the fundamental equation (9), leads to the concept of entropy generation by heat flow, in equation (10).

---

[5] Equation originally unnumbered

[6] That this is the fundamental equation is seen in the “verifications” using this equation, e.g. Miller [14].

[7] The time rate of generation of an extensive quantity is not the time derivative of the quantity, thus $dS/dt \rightarrow \sigma$.

[8] Equation originally unnumbered

Next, *second*, Onsager [7] explicitly defines the concept of entropy flow $J_S$ by heat flow $J_2$ divided by temperature $T$,

$$J_S = \frac{J_2}{T} , \tag{11}$$

in equation (5.7), Onsager [7], p. 421 [subscript 2 for heat flow added],

> “If we write
>
> $$\dot{S}^*(J_{2,n}) = \iint \left( J_{2,n}/T \right) d\Omega \tag{5.7}$$
>
> for the entropy given off to the surroundings […]”

This definition of the concept of entropy flow by heat flow divided by temperature, in equation (11), is *necessary* to make the concept of entropy generation by heat flow, in equation (10), to conform to equation (5.6), Onsager [7], p. 421:[9]

> “The rate of local accumulation of heat equals
>
> $$Tds/dt = -\operatorname{div} J_2 \ [...] \tag{5.6}$$
>
> writing *s* for the local entropy density, […]”

Finally, *third*, with the two concepts of entropy generation by heat flow, in equation (10), and entropy flow by heat flow divided by temperature, in equation (11), Onsager equation (5.6) becomes

$$\frac{ds}{dt} = \frac{-\operatorname{div} J_2}{T} = -\operatorname{div}\left(\frac{J_2}{T}\right) - \frac{1}{T^2} J_2 \operatorname{grad} T = -\operatorname{div} J_S + \sigma_{J_2} , \tag{12}$$

in which the leftmost equality is the equality of Onsager equation (5.6), the middle equality expresses pure mathematical manipulation, and the rightmost equality is based on equations (10) and (11).

Equation (12) contains the two essential concepts of the Onsager-Prigogine entropy model: entropy generation by heat flow and entropy flow by heat flow divided by temperature.

---

[9] Onsager [7] equation (5.6) is used with no derivation nor reference. However, equation (5.6) is obviously based on equation (II), $dS = dQ/T$, of the Clausius entropy model – by expressing the heat term in equation (II) via the divergence of the heat flow.

# 4 EQUIVALENCE

## 4.1 Equivalence of Entropy Accumulation

Consider a physical situation with heat flow **Q**, heat generation $Q_{\mathrm{GEN}}$, and heat absorption $Q_{\mathrm{ABS}}$, the only relevant phenomena.

The entropy balance of a stationary, closed system (control volume, balance volume) according to the Onsager-Prigogine entropy model may be expressed as an extension of equation (13) to include heat generation, e.g. de Groot and Mazur [9], p. 24,

$$\frac{\partial S'''}{\partial t} = -\nabla \cdot \frac{\tilde{\mathbf{Q}}''}{T} - \frac{1}{T^2}\tilde{\mathbf{Q}}'' \cdot \nabla T + \frac{\tilde{Q}'''_{GEN}}{T} . \tag{1}$$

By pure mathematical manipulation, i.e. equation (2), equation (1) may be expressed as

$$\frac{\partial S'''}{\partial t} = \frac{-\nabla \cdot \tilde{\mathbf{Q}}''}{T} + \frac{\tilde{Q}'''_{GEN}}{T} . \tag{13}$$

By the heat balance[10], i.e. equation (3), equation (13) may be expressed as

$$\frac{\partial S'''}{\partial t} = \frac{\tilde{Q}'''_{ABS}}{T} . \tag{14}$$

Because equation (14) must hold true for an arbitrary system, differential or integral, it may be expressed as

$$\frac{dS}{dt} = \frac{\tilde{Q}_{ABS}}{T} , \tag{15}$$

which may be expressed as

$$dS = \frac{dQ}{T} , \tag{II}$$

which is the “second fundamental equation” of the Clausius entropy model.

Now, in any physical situation in which equation (1) hold true for entropy accumulation, equation (II) must hold true for entropy accumulation, which entails that equations (1) and (II) have *the same domain* and makes equations (1) and (II) *equivalent* with respect to the of accumulation entropy – which entails same values of entropy change and same values of entropy, given entropy values at reference state.

This makes, finally, the Clausius entropy model and the Onsager-Prigogine entropy model *equivalent* with respect to entropy accumulation.

---

[10] See appendix A for heat balance.

### 4.2 No Equivalence of All Entropy Phenomena

The Clausius entropy model and the Onsager-Prigogine entropy model are *equivalent* with respect to entropy accumulation – as shown in the comparison of equations (1) and (II), above.

The "second fundamental equation" of the Clausius entropy model, equation (II) may be, equivalently, expressed as equation (14), because equation (II) is a *differential equation*, e.g. Clausius [6], p. 110.

Now, expressing equations (1) and (14) using the shorthand expression of the balance concept, *ACC* = (*IN-OUT*) + *GEN*, gives two equations, the first of Onsager-Prigogine entropy model and the second of Clausius entropy model

$$\underbrace{\frac{\partial S'''}{\partial t}}_{\text{ACC of entropy}} = \underbrace{0}_{\substack{\text{(IN-OUT) of entropy}\\ \text{by bulk matter flow}}} \underbrace{-\nabla\cdot\frac{\tilde{\mathbf{Q}}''}{T}}_{\substack{\text{(IN-OUT) of entropy}\\ \text{by heat flow}}} \underbrace{\underbrace{-\frac{1}{T^2}\tilde{\mathbf{Q}}''\cdot\nabla T}_{\substack{\text{GEN of entropy}\\ \text{by heat flow}}} \underbrace{+\frac{\tilde{Q}'''_{GEN}}{T}}_{\substack{\text{GEN of entropy}\\ \text{by heat generation}}}}_{\text{GEN of entropy}} , \tag{16}$$

$$\underbrace{\frac{\partial S'''}{\partial t}}_{\text{ACC of entropy}} = \underbrace{0}_{\substack{\text{(IN-OUT) of entropy}\\ \text{by bulk matter flow}}} \underbrace{+0}_{\substack{\text{(IN-OUT) of entropy}\\ \text{by heat flow}}} \underbrace{\underbrace{+0}_{\substack{\text{GEN of entropy}\\ \text{by heat flow}}} \underbrace{+\frac{\tilde{Q}'''_{ABS}}{T}}_{\substack{\text{GEN of entropy}\\ \text{by heat absorption}}}}_{\text{GEN of entropy}} . \tag{17}$$

These equations are equivalent with respect to entropy accumulation, but differ in the assumptions of entropy phenomena caused by the heat phenomena – especially, the entropy generations.

### 4.3 Extensions

The scope of the "mechanical theory of heat" of Clausius [1-6] are the *heat and work phenomena* of *closed systems* and this is also the true domain of the "second fundamental equation", equation (II), $dS = dQ/T$, of the Clausius entropy model – because "reversibility" is false. This domain leads to equations (16) and (17) of the Onsager-Prigogine entropy model and the Clausius entropy model, respectively.

The extensions of equations (16) and (17) to open systems and to other (than heat and work) phenomena are

$$\underbrace{\frac{\partial S'''}{\partial t}}_{\text{ACC of entropy}} = \underbrace{-\nabla\cdot S'''\tilde{\mathbf{u}}_{bulk}}_{\substack{\text{(IN-OUT) of entropy}\\ \text{by bulk matter flow}}} \underbrace{-\nabla\cdot\frac{\tilde{\mathbf{Q}}''}{T}}_{\substack{\text{(IN-OUT) of entropy}\\ \text{by heat flow}}} \underbrace{\underbrace{-\frac{1}{T^2}\tilde{\mathbf{Q}}''\cdot\nabla T}_{\substack{\text{GEN of entropy}\\ \text{by heat flow}}} \underbrace{+\frac{\tilde{Q}'''_{GEN}}{T}}_{\substack{\text{GEN of entropy}\\ \text{by heat generation}}} \underbrace{+\tilde{S}'''_{GEN,O}}_{\substack{\text{GEN of entropy}\\ \text{by other phenomena}}}}_{\text{GEN of entropy}} , \tag{18}$$

$$\underbrace{\frac{\partial S'''}{\partial t}}_{\text{ACC of entropy}} = \underbrace{-\nabla\cdot S'''\tilde{\mathbf{u}}_{bulk}}_{\substack{\text{(IN-OUT) of entropy}\\ \text{by bulk matter flow}}} \underbrace{+0}_{\substack{\text{(IN-OUT) of entropy}\\ \text{by heat flow}}} \underbrace{\underbrace{+0}_{\substack{\text{GEN of entropy}\\ \text{by heat flow}}} \underbrace{+\frac{\tilde{Q}'''_{ABS}}{T}}_{\substack{\text{GEN of entropy}\\ \text{by heat absorption}}} \underbrace{+\tilde{S}'''_{GEN,O}}_{\substack{\text{GEN of entropy}\\ \text{by other phenomena}}}}_{\text{GEN of entropy}} . \tag{19}$$

These equations are equivalent with respect to entropy accumulation, but entail different entropy generations.

# 5 JUSTIFICATION AND FALSIFICATION

## 5.1 Equivalence

The equivalence of the Clausius and Onsager-Prigogine entropy models is justified by pure mathematics and the balance of heat, i.e. equations (2) and (3),

$$\nabla \cdot \frac{\tilde{\mathbf{Q}}''}{T} = \frac{\nabla \cdot \tilde{\mathbf{Q}}''}{T} - \frac{\tilde{\mathbf{Q}}'' \cdot \nabla T}{T^2} , \tag{2}$$

$$\tilde{Q}'''_{ABS} = -\nabla \cdot \tilde{\mathbf{Q}}'' + \tilde{Q}'''_{GEN} , \tag{3}$$

that entail that the equations of the Clausius and Onsager-Prigogine entropy models – e.g. equation (II), $dS = dQ/T$, of the Clausius entropy model – are *equivalent with respect entropy accumulation*.

However, according to Clausius [1-6], equation (II) holds true for “reversible” phenomena, *only*, while the equations of the Onsager-Prigogine entropy model are taken to be not thus restricted, which entails that the equivalence and “reversibility” are contradictory and, thus, not both true.

## 5.2 “Reversibility”

According to Clausius [1-6], equation (II), $dS = dQ/T$, holds true for “reversible” phenomena and for “reversible” phenomena *only*.

Accordingly, the essence of “reversibility” is equation (II), $dS = dQ/T$, *not holding true for not “reversible” phenomena*. [11]

### 5.2.1 Justification

“Reversibility” is justified by the claims by Clausius [1-6], only.

The truth of “reversibility” would be corroborated by evidence for equation (II), $dS = dQ/T$, *not holding true for not “reversible” phenomena*. [12]

No such evidence has been presented – neither by Clausius [1-6] nor latter entropy modeling studies.

### 5.2.2 Falsification

“Reversibility” is falsified by the contradiction with *entropy history independence*.

The falsity of “reversibility” is corroborated by evidence for equation (II), $dS = dQ/T$, *holding true for not “reversible” phenomena*.

There is ample evidence for equation (II), $dS = dQ/T$, *holding true for not “reversible” phenomena*, see **2.5 Falsification of “Reversibility”** and appendices D and E.

---

[11] The essence of “reversibility” is not equation (II), $dS = dQ/T$, holding true for “reversible” phenomena.

[12] “Reversibility” is not corroborated by evidence for equation (II), $dS = dQ/T$, holding true for “reversible” phenomena.

## 6 SUMMARY

**1.**

There are currently two – main, continuum – models of entropy: the (“reversible”, “equilibrium”) Clausius entropy model and (“irreversible”, “nonequilibrium”) Onsager-Prigogine entropy model.

The analysis demonstrates that equation (II), $dS = dQ/T$, of the Clausius entropy model,

$$dS = \frac{dQ}{T}\,, \tag{II}$$

and the Onsager-Prigogine entropy model, for the same physical situation,

$$\frac{\partial S'''}{\partial t} = -\nabla \cdot \frac{\tilde{\mathbf{Q}}''}{T} - \frac{1}{T^2}\tilde{\mathbf{Q}}'' \cdot \nabla T + \frac{\tilde{Q}'''_{GEN}}{T}\,, \tag{1}$$

have the *same domain* and are *equivalent* with respect to the accumulation of entropy.

The demonstration is based on pure mathematical manipulation and the balance of heat.

**2.**

The essence of the Clausius entropy model is that equation (II), $dS = dQ/T$, holds true for “reversible” phenomena and for “reversible” phenomena, *only*, which may be expressed as

$$dS_{\text{"rev"}} = \frac{dQ_{\text{"rev"}}}{T}\,, \tag{5}$$

$$dS_{\text{not "rev"}} \neq \frac{dQ_{\text{not "rev"}}}{T}\,. \tag{6}$$

This, however, with $dQ_{\text{"rev"}} = dQ_{\text{not "rev"}}$, leads to a contradiction with *entropy history independence*, which, if true, falsifies “reversibility”.

**3.**

That “reversibility” is false is corroborated by experimental evidence from engineering calculations using equation (II), consistently found to conform to the physical reality, and by conceptual and numeric examples – also for cycles – using equation (II) for not “reversible” phenomena.

**4.**

The results – of falsity of “reversibility” and of equivalence of the two entropy models – would be falsified by *a single evidence* for equation (II) *not holding true for not “reversible” phenomena*.

**5.**

The main consequence of the equivalence of the two entropy models is a signification simplification of the language used in the context of entropy modeling – which cannot fail to enhance understanding of entropy modeling.

# APPENDIX A: BALANCE

## A.1 Balance Proposition

The balance proposition may be taken as an axiom, as in Reynolds [15], p. 9:

> “AXIOM I: Any change whatsoever in the quantity of any entity within a closed surface can only be effected in one or other of the two distinct ways: (1) it may effected by the production or destruction of the entity within the surface, or (2) by the passage of the entity across the surface.”

Thus, for simplicity, it is taken as an axiom that

*For any extensive quantity, for any system, and for any instant or interval of time, accumulation equals net input plus net generation.* (A.1)

The balance proposition may be expressed as[13]

$$ACC = \left(IN - OUT\right) + GEN \,, \tag{A.2}$$

where the terms have meaning with respect to a system C and its boundary B, as in figure A.1, only.

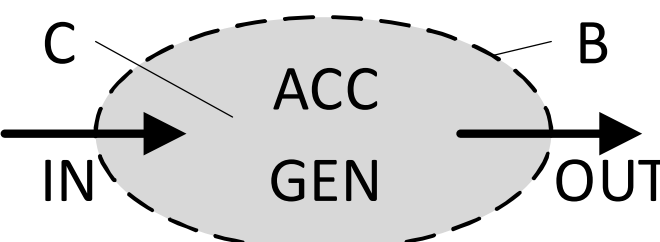

**Figure A.1 The system (i.e. control volume, balance volume) and the terms of the balance**

## A.2 “Transit”, “Transport” and “Transfer”

The terms “transit”, “transport”, and “transfer” may are used with varying meanings.

Here, the terms refer to the following dichotomy of the flux of an extensive quantity

$$X_{transit} = X_{transport} + X_{transfer} \,, \tag{A.3}$$

in which “transit” refers to the *total flux*, “transport” refers to the *flux due to flux of bulk matter*, and “transfer” refers to the *flux with respect to (flux of) bulk matter*.

## A.3 Balance

For an open, stationary system C with boundary B, the balance of the extensive quantity X may be expressed, using the dichotomy of “transport” and “transfer” for the net input term, as[14]

$$\frac{d}{dt}\int_C X'''dV = -\int_B \mathbf{n}_B \cdot X'''\tilde{\mathbf{u}}_{bulk} dA - \int_B \mathbf{n}_B \cdot \tilde{\mathbf{X}}''_{transfer} dA + \int_C \tilde{X}'''_{GEN} dV \,, \tag{A.4}$$

from which, using the divergence theorem,

$$\frac{\partial X'''}{\partial t} = -\nabla \cdot X'''\tilde{\mathbf{u}}_{bulk} - \nabla \cdot \tilde{\mathbf{X}}''_{transfer} + \tilde{X}'''_{GEN} \,. \tag{A.5}$$

---

[13] More explicitly, ACC = $IN_{net}$ + $GEN_{net}$ or ACC = (IN-OUT) + (GEN-DES).

[14] A tilde denotes a time rate of a quantity and primes denote a quantity per area or per volume.

## A.4 Example

The generic balance of moles of component i of matter for an open, stationary system, based on equation (A.5), is

$$\underbrace{\frac{\partial\left(x_i n'''\right)}{\partial t}}_{\text{ACC}} = \underbrace{-\nabla\cdot\left(x_i n'''\right)\tilde{\mathbf{u}}_{bulk}}_{(\text{IN-OUT})_{transport}} \quad \underbrace{-\nabla\cdot\tilde{\mathbf{n}}''_{i,transfer}}_{(\text{IN-OUT})_{transfer}} \quad \underbrace{+\tilde{n}'''_{GEN}}_{\text{GEN}} \; . \tag{A.6}$$

in which $x_i$ is the mole fraction of component i, $n'''$ is the bulk molar concentration, the rate of transfer may be calculated, e.g., as

$$\tilde{\mathbf{n}}''_{i,transfer} = -\nabla\cdot\left(D_i x_i\right), \tag{A.7}$$

and the rate of generation may be calculated, e.g., as

$$\tilde{n}'''_{i,GEN} = \sum_k \tilde{r}'''_{i,k} = \sum_k \upsilon_{i,k}\tilde{r}'''_k , \tag{A.8}$$

## A.6 Heat

Heat may be considered a *special, extensive, bookkeeping quantity* of energy

By definition, heat $Q$, *as such*, does not exist (has zero value anywhere anytime), while *heat flow* (or “heat”) $\mathbf{Q}$ is energy in transit due to temperature difference, *heat generation* $Q_{GEN}$ is interconversion of a kind of energy to heat, and *heat absorption* $Q_{ABS}$ is interconversion of heat to internal energy.

The generic balance of heat for an open, stationary system, based on equation (A.5), is [15]

$$\underbrace{\frac{\partial Q'''}{\partial t}}_{\text{ACC}} = \underbrace{-\nabla\cdot Q'''\tilde{\mathbf{u}}_{bulk}}_{(\text{IN-OUT})_{transport}} \quad \underbrace{-\nabla\cdot\tilde{\mathbf{Q}}''}_{(\text{IN-OUT})_{transfer}} \quad \underbrace{+\left(\tilde{Q}'''_{GEN} - \tilde{Q}'''_{ABS}\right)}_{\text{GEN}} \; . \tag{A.9}$$

in which $Q'''$ is heat per volume. Because, by definition, $Q''' \equiv 0$, this leads to

$$\tilde{Q}'''_{ABS} = -\nabla\cdot\tilde{\mathbf{Q}}'' + \tilde{Q}'''_{GEN} , \tag{A.10}$$

which may be expressed, equivalently,

$$dQ_{ABS} = dQ_{(IN-OUT)} + dQ_{GEN} . \tag{A.11}$$

The special attributes of heat has led to special notations to be used in the heat term of the equation (II), $dS = dQ/T$. An alternative is to express equation (II) as an ordinary differential equation as

$$\frac{dS}{dt} = \frac{\tilde{Q}}{T}, \quad \text{or} \quad \frac{\partial S'''}{\partial t} = \frac{\tilde{Q}'''}{T} , \tag{A.12}$$

in which the numerators in the RHSs are time rates of heat (transferred, generated, or absorbed), see e.g. Van Wylen and Sonntag [16], p. 107, 219.

---

[15] Heat generation, $Q_{GEN}$, is non-negative, while the balance quantity net generation, GEN, is not thus restricted. This ambiguity could be avoided by expressing the balance as ACC = (IN-OUT) + (GEN-DES).

## A.4 Energy

The generic balance of energy for an open, stationary system, based on equation (A.5), is

$$\frac{\partial E'''}{\partial t} = -\nabla \cdot E'''\tilde{\mathbf{u}}_{bulk} - \nabla \cdot \tilde{\mathbf{E}}''_{transfer} + \tilde{E}'''_{GEN} . \tag{A.13}$$

First, take energy conservation, i.e. energy no-generation anywhere anytime, and express this as

$$E_{GEN} \equiv 0 . \tag{A.14}$$

Second, take energy transfer equal to heat transfer plus work transfer and express this as[16]

$$\tilde{\mathbf{E}}''_{transfer} = \tilde{\mathbf{Q}}'' + \tilde{\mathbf{W}}'' . \tag{A.15}$$

Third, take a closed system, the definition of which excludes the flux of matter through the boundary and thus transport of any quantity, e.g. energy, with matter, i.e.

$$\tilde{\mathbf{E}}''_{transport} = E'''\tilde{\mathbf{u}}_{bulk} = 0 . \tag{A.16}$$

The insertion of equations (A.14-A.16) into equation (A.13) gives

$$\frac{\partial E'''}{\partial t} = -\nabla \cdot \tilde{\mathbf{Q}}'' - \nabla \cdot \tilde{\mathbf{W}}'' , \tag{A.17}$$

which may be expresses as

$$\frac{dE}{dt} = \tilde{Q}_{(IN-OUT)} + \tilde{W}_{(IN-OUT)} , \tag{A.18}$$

and as

$$dE = dQ + dW . \tag{A.19}$$

## A.5 Entropy

The generic balance of entropy for an open, stationary system, based on equation (A.5), is

$$\frac{\partial S'''}{\partial t} = -\nabla \cdot S'''\tilde{\mathbf{u}}_{bulk} - \nabla \cdot \tilde{\mathbf{S}}''_{transfer} + \tilde{S}'''_{GEN} . \tag{A.20}$$

For entropy, the balance quantities depend on the entropy model used, whether the Clausius entropy model or the Onsager-Prigogine entropy model, see 4 Equivalence, and, thus, these are not considered, here.

---

[16] Because heat and work have *zero value anywhere anytime*, there is neither heat nor work *transport*. Thus, to simplify the equations, the *transfer* of heat and work is not denoted, i.e. $\mathbf{Q} = \mathbf{Q}_{\text{transfer}}$, $\mathbf{W} = \mathbf{W}_{\text{transfer}}$, $Q_{(\text{IN-OUT})} = Q_{(\text{IN-OUT}),\text{transfer}}$, and $W_{(\text{IN-OUT})} = W_{(\text{IN-OUT}),\text{transfer}}$.

## APPENDIX B: CLAUSIUS ON TEMPERATURE DIFFERENCE

According to Clausius [4], p. 133, [6], p. 212, phenomena with non-zero temperature difference are not “reversible”.

The claim *may be* based on Clausius, analogously to Carnot, considering heat engines with the assumption of “reservoirs” with temperatures equal to the temperatures of the working fluid during the isothermal steps.

However, the temperatures of the “reservoirs” are irrelevant, according to Clausius:

Clausius [2], p. 501, [underline added]:

> “In diesem Falle ist es natürlich einerlei, ob die in der Gleichung (II) verkommende Gröfse die Temperatur des eben benutzten Wärmereservoirs oder die augenblickliche Temperatur des veränderlichen Körpers darstellt, da beide gleich sind. Hat man aber einmal für t die letztere Bedeutung eingeführt, so ist leicht zu sehen, dafs man nun den Wärmereservoiren beliebige andere Temperaturen beilegen kann, ohne dafs dadurch der Ausdruck $\int \frac{dQ}{T}$ irgend eine Aenderung erleidet, welche die Gültigkeit der vorigen Gleichung beeinträchtigen könnte. Da bei dieser Bedeutung von t die einzelnen Wärmereservoire nicht mehr besonders berücksichtigt zu werden brauchen, so pflegt man auch die Wärmemengen nicht auf sie, sondern auf den veränderlichen Körper zu beziehen, indem man angiebt, welche Wärmemengen der Körper während seiner Veränderungen nach einander aufnimmt oder abgiebt.“

Clausius [4], p. 130, [underline added]:

> “In this case it is of course of no importance whether t, in the equation (II), represents the temperature of the reservoir of heat just employed, or the momentary temperature of the changing body, inasmuch as both are equal. The latter signification being once adopted, however, it is easy to see that any other temperatures may be attributed to the reservoirs of heat without producing thereby any change in the expression $\int \frac{dQ}{T}$ which shall be prejudicial to the validity of the foregoing equation. As with this signification of t the several reservoirs of heat need no longer enter into consideration, it is customary to refer the quantities of heat, not to them, but to the changing body itself, by stating what quantities of heat this body successively receives or imparts during its modifications.”

Clausius [5], p. 111, [underline added]:

> “In diesem Falle ist es natürlich einerlei, ob man die Temperatur einer übergehenden Wärmemenge der Temperatur des Wärmereservoirs oder der augenblicklichen Temperatur des veränderlichen Körpers gleichsetzen will, da beide unter einander übereinstimmen. Hat man aber einmal die letztere Wahl getroffen, und festgesetzt, dass bei der Bildung der Gleichung(VII.) für jedes Wärmeelement dQ diejenige Temperatur in Rechnung gebracht werden soll, welche der veränderliche Körper bei seiner Aufnahme gerade hat, so kann man nun den Wärmereservoiren auch beliebige andere Temperaturen zuschreiben, ohne dass dadurch der Ausdruck $\int \frac{dQ}{\tau}$ eine Aenderung erleidet.“

Clausius [6], p. 106, [underline added]:

> “In this case it is obviously the same thing whether we consider the temperature of a quantity of heat which is being transferred as being equal to that of the reservoir or of the variable body, since these are practically the same. If however we choose the latter and suppose that in forming Equation VII. every element of heat dQ is taken of that temperature which the variable body possesses at the moment it is taken in, then we can now ascribe to the heat reservoirs any other temperatures we please, without thereby making any alteration in the expression $\int \frac{dQ}{\tau}$.”

## APPENDIX C: "REVERSIBILITY" IN TEXTBOOKS

According to Clausius [4], p. 133, [6], p. 212, phenomena with *temperature difference* and *heat generation* are not "reversible" and are excluded from the domain of the "second fundamental equation" (II), $dS = dQ/T$.

Zemansky [17], p. 179 [underline added]:

> "**10.7. Entropy and Irreversibility**. When a system undergoes an irreversible process between an initial equilibrium state and a final equilibrium state, the entropy change of the system is equal to
>
> $$S_f - S_i = \int_R \frac{dQ}{T}$$
>
> where R indicates any reversible process arbitrarily chosen by which the system may be brought from the given initial state to the given final state. No integration is performed over the original irreversible path. The irreversible process is replaced by a reversible one."

Kondepudi and Prigogine [18], pp. 84-85 [underline added]:

> "The usefulness of the concept of entropy and the Second Law depends on our ability to define entropy of a physical system in a calculable way. Using (3.3.3), if the entropy $S_0$ of a reference or standard state is defined, then the entropy of an arbitrary state $S_X$ can be obtained through a reversible process that transforms the state 0 to the state X (Fig. 3.6).
>
> $$S_X - S_0 = \int_0^X \frac{dQ}{T} \tag{3.4.1}$$
>
> (In practice $dQ$ is measured with the knowledge of the heat capacity using $dQ = CdT$.) In a real system the transformation from state 0 to state X occurs in a finite time and involves irreversible processes along the path I. In classical thermodynamics it is assumed that every irreversible transformation that occurs in nature can also be achieved through a reversible process for which (3.4.1) is valid."

The essence of both excerpts is that equation (II), $dS = dQ/T$, *may be used* for the calculation of the entropy change of a system by an "irreversible" process – by assuming the "irreversible" process replaced by an imaginary "reversible" process (with the same amount of heat, obviously), which scheme is assumed to be possible for every process "in nature". [17]

Now, if equation (II), $dS = dQ/T$, *may be used* for the calculation of not "reversible" phenomena, then equation (II), $dS = dQ/T$, *holds true* for not "reversible" phenomena – in any adequate meaning of "to hold true" (in this context).

Thus, both excerpts are in contradiction with equation (II), $dS = dQ/T$, holding true for "reversible" phenomena, *only*, which is maintained by Clausius [1-6] and, also, by Zemansky [17], p. 174, and Kondepudi and Prigogine [18], p. 82.

This entails that equation (II), $dS = dQ/T$, holds true for not "reversible" phenomena, also.

---

[17] The excerpts entail that $dS = dQ_{\text{not "rev"}}/T = dQ_{\text{"rev"}}/T = dQ/T$, given $dQ_{\text{not "rev"}} = dQ_{\text{"rev"}} = dQ$ – which is necessary, if entropy is history independent ("path independent", "state function"), and entails that "reversibility" is false.

## APPENDIX D: EXAMPLES OF EQUATION (II) IN ENGINEERING PRACTICE

According to Clausius [4], p. 133, [6], p. 212, phenomena with *temperature difference* and *heat generation* are not "reversible" and are excluded from the domain of the "second fundamental equation" (II), $dS = dQ/T$.

### D.1 Heating with Temperature Difference

First, consider *steady* (state), i.e. time independent, isobaric, lossless heating of pure nitrogen from $p_1$ = 1.0 bar, $T_1$ = 0.0 C to $T_2$ = 100 C with high flow of 3 bar saturated steam (T = 134 C).

The case calculated by Aspen Plus®, of Aspen Technology, Inc., is presented in figure D.1.

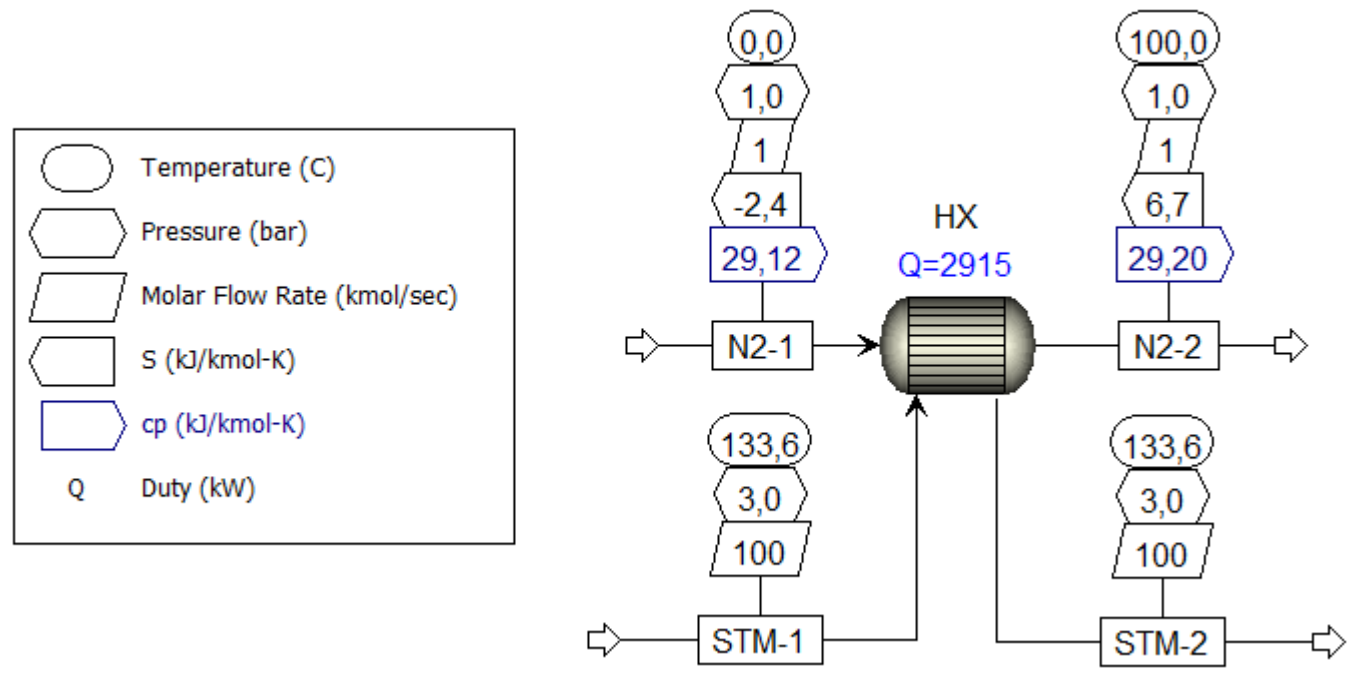

**Figure D.1: Heating with temperature difference**

Second, consider *batch*, i.e. time dependent, isobaric, lossless heating of an ideal gas with $C_p$ = 29.16 J/K/mol within a closed system from $p_1$ = 1.0 bar, $T_1$ = 0.0 C and $S_1$ = -2.4 J/K/mol to $T_2$ = 100 C.

The entropy modeling, using the "second fundamental equation" (II), $dS = dQ/T$, gives

$$dS = \int_{T_1}^{T_2} \frac{dQ}{T} = \int_{T_1}^{T_2} \frac{dQ_{IN}}{T} = \int_{T_1}^{T_2} \frac{dQ_{ABS}}{T} = \int_{T_1}^{T_2} \frac{C_p dT}{T} = C_p \ln\left(\frac{T_2}{T_1}\right) = 9{,}1 \text{ J/K/mol} \tag{D.1}$$

$$S_2 = S_1 + dS = (-2{,}4 + 9{,}1) \text{ J/K/mol} = 6{,}7 \text{ J/K/mol} \tag{D.2}$$

Now, it is seen that the result for $S_2$ is the same as calculated by Aspen Plus®, in figure D.1.

Thus, the numerical results obtained using the "second fundamental equation" (II), $dS = dQ/T$, equal the results of engineering practice for this case of heating with *temperature difference* explicitly excluded from the domain of the equation (II), Clausius [4], p. 133, [6], p. 212.

This entails that equation (II), $dS = dQ/T$, holds true for not "reversible" phenomena, also.

## D.2 Compression with Heat Generation

First, consider *steady* (state), i.e. time independent, adiabatic, ideal (i.e. lossless) compression of pure nitrogen from $p_1$ = 1.0 bar, $T_1$ = 0.0 C to $p_2$ = 3.0 bar and the same with losses with efficiency of 0.5.

The cases calculated by Aspen Plus®, of Aspen Technology, Inc., are presented in figure D.2.

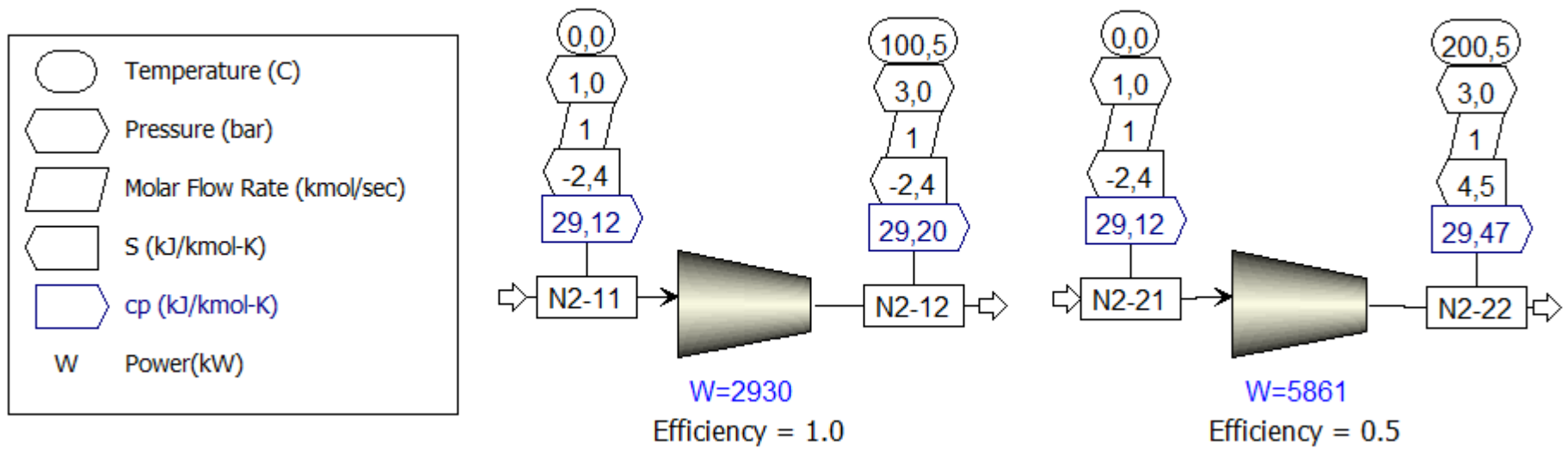

**Figure D.2: Compression with heat generation**

Second, consider *batch*, i.e. time dependent, adiabatic, lossless compression of an ideal gas with $C_p$ = 29.16 J/K/mol within a closed system from $p_1$ = 1.0 bar, $T_1$ = 0.0 C and $S_1$ = -2.4 J/K/mol to $p_2$ = 3.0 bar and the same with losses with efficiency of 0.5.

The heat generation, equal to the work input “lost”, is

$$dQ_{GEN} = dW_{IN,lost} = \frac{1-\eta}{\eta} dW_{IN,ideal} = \frac{1-\eta}{\eta} \frac{RT_1}{k-1} \left[ \left( \frac{p_2}{p_1} \right)^{\left( \frac{k-1}{k} \right)} - 1 \right]. \tag{D.3}$$

The entropy modeling, using the “second fundamental equation” (II), $dS = dQ/T$, gives

$$dS = \int \frac{dQ}{T} = \int \frac{dQ_{ABS}}{T} = \int \frac{dQ_{GEN}}{T} = \int_{T_{2,ideal}}^{T_2} \frac{C_p dT}{T} = C_p \ln \left( \frac{T_2}{T_{2,ideal}} \right). \tag{D.4}$$

The results are presented in Table D.1.

| efficiency | 1,0 | 0,5 | - |
|---:|---:|---:|---|
| dW,IN | 2 094,6 | 4 189,2 | J/mol |
| dQ,IN | 0,0 | 0,0 | J/mol |
| dW,lost | 0,0 | 2 094,6 | J/mol |
| dQ,GEN | 0,0 | 2 094,6 | J/mol |
| dQ,ABS | 0,0 | 2 094,6 | J/mol |
| dS | 0,0 | 6,9 | J/mol |
| T,2 | 100,5 | 201,0 | C |
| S,2 | - 2,4 | 4,5 | J/K/mol |

Input specifications grayed.

**Table D.1.**

Now, it is seen that the results for $T_2$ and $S_2$ are the same as calculated by Aspen Plus®, in figure D.2.

Thus, the numerical results obtained using the “second fundamental equation” (II), $dS = dQ/T$, equal the results of engineering practice for this case of compression with *heat generation* explicitly excluded from the domain of the equation (II), Clausius [4], p. 133, [6], p. 212.

This entails that equation (II), $dS = dQ/T$, holds true for not “reversible” phenomena, also.

# APPENDIX E: EXAMPLES OF USE OF EQUATION (II)

## E.1 Heating

Consider the heating of a solid cube of one kilogram of silver from initial, spatially uniform, temporally constant temperature $T_1 = 0$ C to final, spatially uniform, temporally constant temperature $T_2 = 100$ C, as in figure E.1. Assume constant physical properties and take that s(0 C) = 0 J/K/kg.

In case a) by increasing, with zero time rate, the temperature of one side of the body from $T_1$ to $T_2$.

In case b) by fixing one side of the body to constant temperature $T_2$.

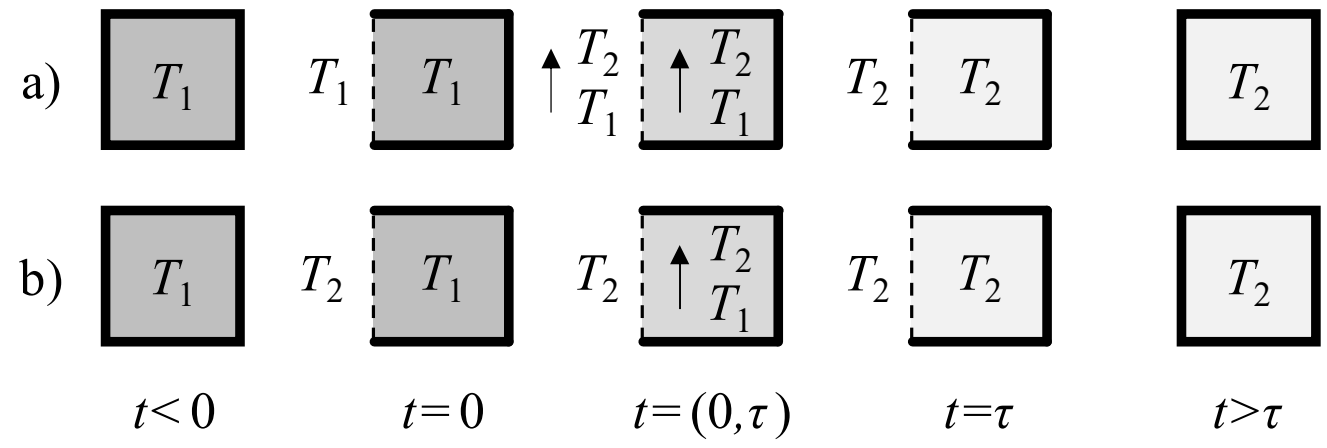

**Figure E.1: Heating of a body**

Because in case a) the heat transfer is with zero temperature difference, the phenomenon is “reversible” and equation (II), $dS = dQ/T$, holds true for case a) – according to Clausius [1-6]

Because in case b) the heat transfer is with non-zero temperature difference, the phenomenon is not “reversible” and equation (II), $dS = dQ/T$, does not hold true for case b) – according to Clausius [1-6].

**Case a):** In case a) the temperature of the system is uniform, because of the zero time rate of heating. Now, the entropy change of the uniform temperature system in the case a), using equation (II), $dS = dQ/T$, is

$$\Delta S_a = \int_{T_1}^{T_2} \frac{dQ_{ABS}}{T} = \int_{T_1}^{T_2} \frac{mc_p dT}{T} = mc_p \ln\frac{T_2}{T_1} = 1 \cdot 240 \cdot \ln\frac{(100+273)}{(0+273)} \text{ J/K} = 75 \text{ J/K}\,, \qquad \text{(E.1)}$$

and thus the final specific entropy $s_{2,a}$ is

$$s_{2,a} = s_1 + \frac{\Delta S_a}{m} = \left(0 + \frac{75}{1}\right) \text{J/K/kg} = 75 \text{ J/K/kg}\,. \qquad \text{(E.2)}$$

**Case b.1) - Uniform temperature:** In case b), *first*, assume *uniform temperature* during the heating.

Now, the entropy change of the uniform temperature system in the case b), using equation (II), $dS = dQ/T$, is

$$\Delta S_b = \int_{T_1}^{T_2} \frac{dQ_{ABS}}{T} = \int_{T_1}^{T_2} \frac{mc_p dT}{T} = mc_p \ln\frac{T_2}{T_1} = 1 \cdot 240 \cdot \ln\frac{(100+273)}{(0+273)} \text{ J/K} = 75 \text{ J/K}\,, \qquad \text{(E.3)}$$

and thus the final specific entropy $s_{2,b}$ is

$$s_{2,b} = s_1 + \frac{\Delta S_b}{m} = \left(0 + \frac{75}{1}\right) \text{J/K/kg} = 75 \text{ J/K/kg}\,. \qquad \text{(E.4)}$$

Thus, the entropy change $\Delta S$ and the final specific entropy $s$ have the same magnitude in cases a) and b).

This entails that equation (II), $dS = dQ/T$, holds true for not “reversible” phenomena, also.

**Case b.2) – Non-uniform temperature:** In case b), *second*, do not assume uniform temperature during the heating.

Thus, the temperature $T$ and specific entropy $s$ during the heating are calculated from the *field equations*, Truesdell and Toupin [13], p. 226-233,

$$\rho c_p \frac{\partial T}{\partial t} = -\nabla \cdot \tilde{\mathbf{Q}}'', \tag{E.5}$$

$$\rho \frac{\partial s}{\partial t} = \frac{-\nabla \cdot \tilde{\mathbf{Q}}''}{T}. \tag{E.6}$$

Equations (E.5) and (E.6) are based on the equations (I) and (II), respectively, assuming i) a differential, closed system, ii) (internal) energy a function of temperature only, iii) no work phenomena, and iv) no heat generation.

For the numerical solution, equations (E.5) and (E.6) are discretized and solved for one spatial dimension, i.e., for $T(z,t)$ and $s(z,t)$, respectively, with $\Delta z = z_{\mathrm{cube}}/100 = 0.5$ mm and $\Delta t = 10^{-5}$ s and

$$T(z,t) = T_1; \quad \forall z, t < 0, \tag{E.7}$$

$$T(z,t) = T_2; \quad z = 0, t \geq 0. \tag{E.8}$$

The results are presented in figures E.2 and E.3 as temperature $T$ and specific entropy $s$ profiles with respect to the dimensionless linear dimension $Z$ and time $t$.

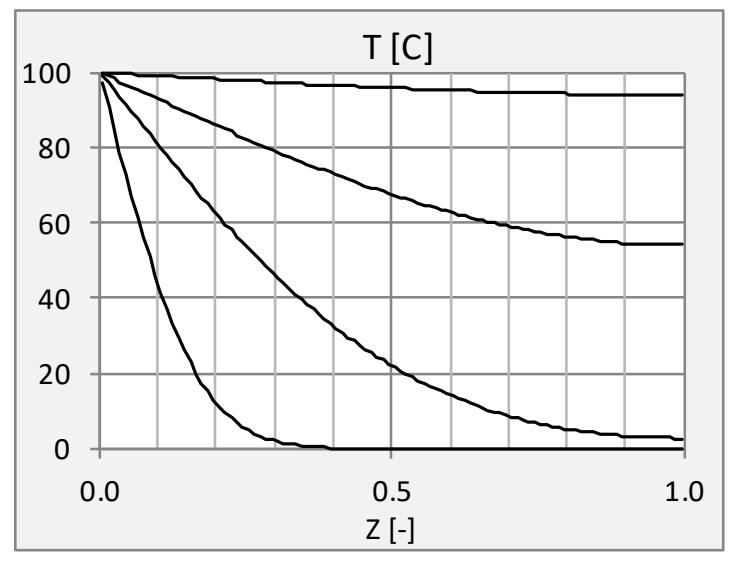

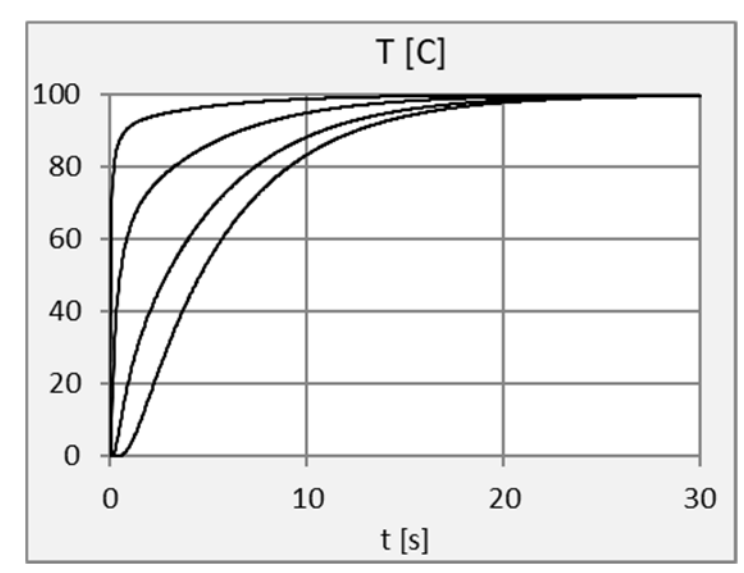

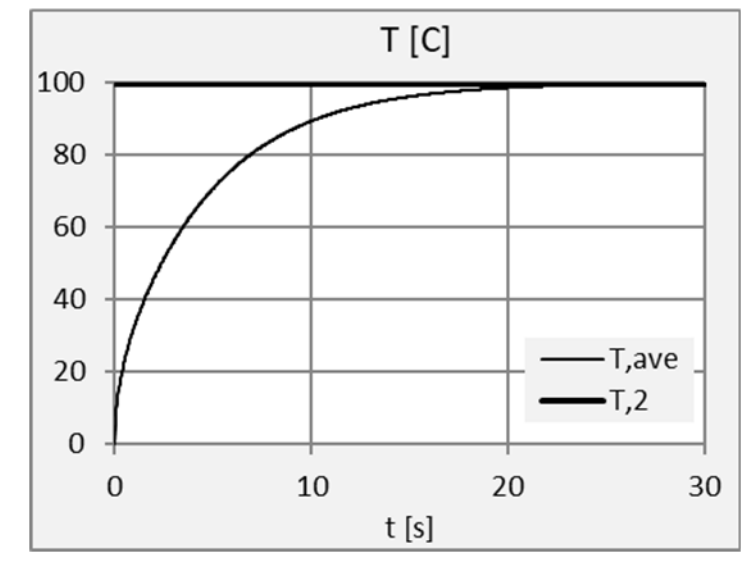

**Figure E.2: Temperature Profiles**

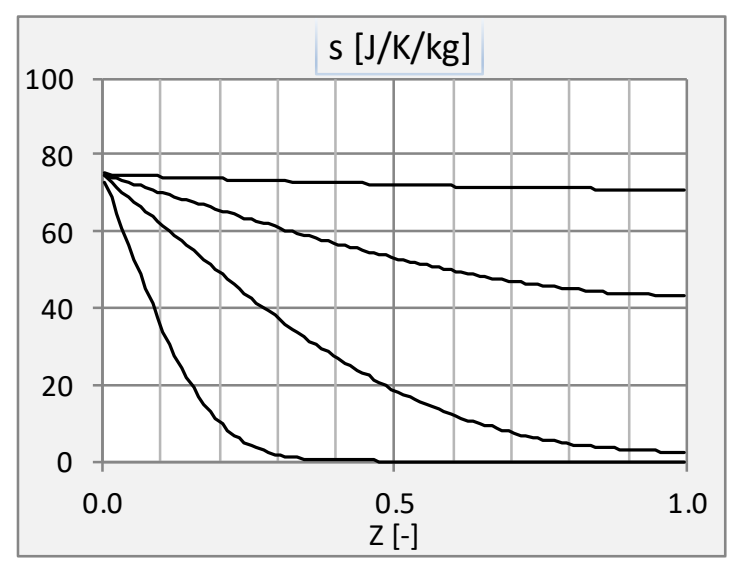

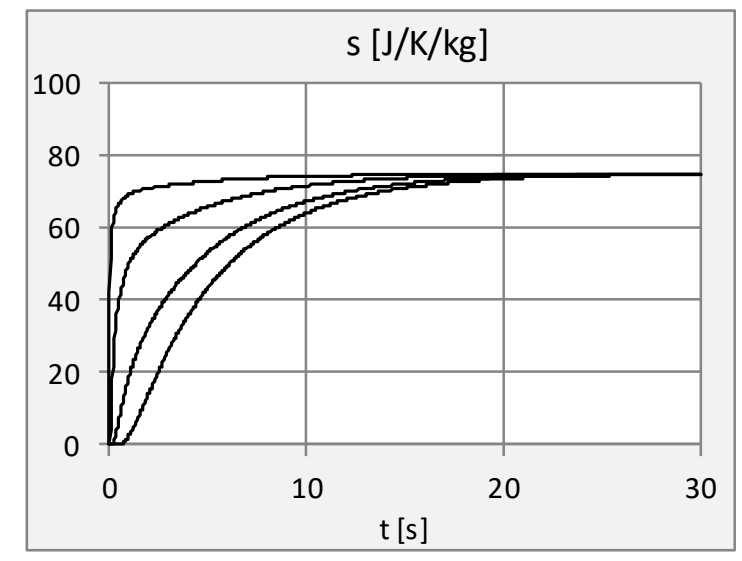

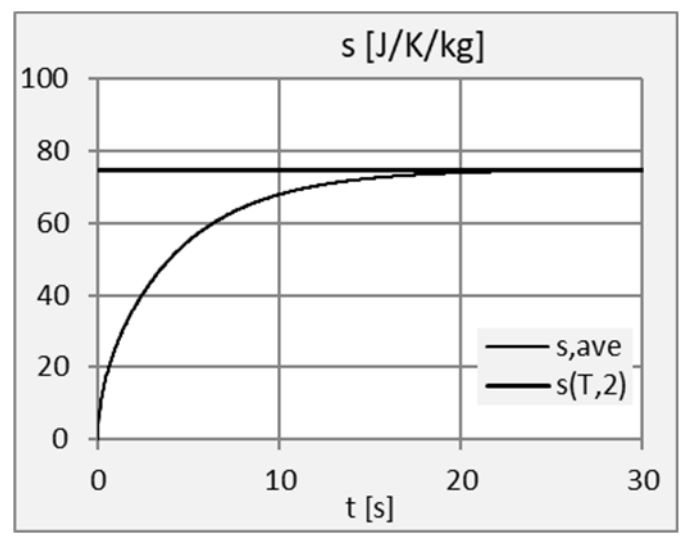

**Figure E.3: Specific Entropy Profiles**

It is seen that, when the average temperature $T_{\mathrm{ave}}$ approaches $T_2 = 100$ C, the average specific entropy $s_{\mathrm{ave}}$ approaches $s = 75$ J/K/kg, which equals the result calculated by the “reversible” model.

This entails that equation (II), $dS = dQ/T$, holds true for not “reversible” phenomena, also.

## E.2 Isothermal Expansion

Consider the isothermal expansion of an ideal gas within a closed system from an initial, spatially uniform, temporally constant state ($p_1$, $V_1$, $T_1$) to a final, spatially uniform, temporally constant state ($p_2$, $V_2$, $T_1$), as in the cases in figure E.4.

In case a) with heat transfer with zero temperature difference, in case b) with heat transfer with non-zero temperature difference, and in cases c) and d) with heat generation due to work transfer turning a mixer with zero time rate and non-zero time rate, respectively.

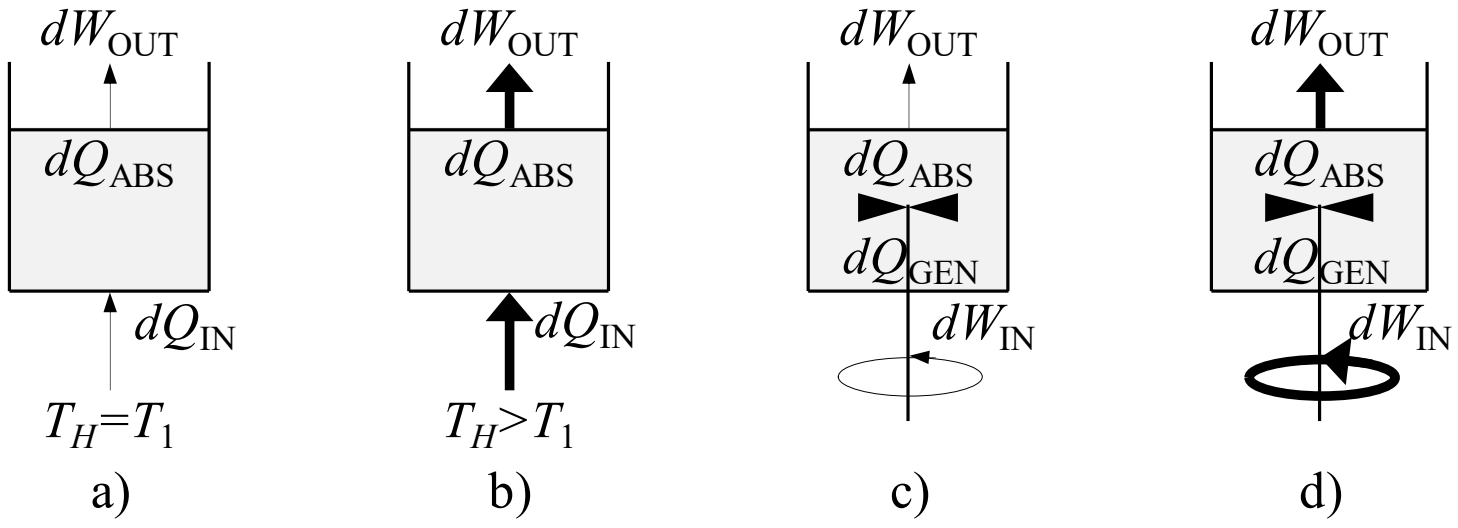

**Figure E.4: Isothermal expansion.**

Because in case a) the heat transfer is with zero temperature difference, the phenomenon is “reversible” and equation (II), $dS = dQ/T$, holds true for case a) – according to Clausius.

Because and in case b) the heat transfer is with non-zero temperature difference and in cases c) and d) the heat absorbed is heat generated, the phenomena are not “reversible” and equation (II), $dS = dQ/T$, does not hold true for cases b), c), and d) – according to Clausius.

For simplicity, assume uniform temperature and pressure.[18] Because the uniform temperature is constant at $T_1$, the integrations using equation (II), $dS = dQ/T$, for the four cases give

$$\Delta S_a = \int_0^Q \frac{dQ_{ABS,a}}{T_1} = \int_0^Q \frac{dQ_{IN,a}}{T_1} = \frac{Q_{IN,a}}{T_1} \tag{E.9}$$

$$\Delta S_b = \int_0^Q \frac{dQ_{ABS,b}}{T_1} = \int_0^Q \frac{dQ_{IN,b}}{T_1} = \frac{Q_{IN,b}}{T_1} \tag{E.10}$$

$$\Delta S_c = \int_0^Q \frac{dQ_{ABS,c}}{T_1} = \int_0^Q \frac{dQ_{GEN,c}}{T_1} = \int_0^W \frac{dW_{IN,c}}{T_1} = \frac{W_{IN,c}}{T_1} \tag{E.11}$$

$$\Delta S_d = \int_0^Q \frac{dQ_{ABS,d}}{T_1} = \int_0^Q \frac{dQ_{GEN,d}}{T_1} = \int_0^W \frac{dW_{IN,d}}{T_1} = \frac{W_{IN,d}}{T_1} \tag{E.12}$$

For isothermal expansion of an ideal gas between the initial and final, spatially uniform, temporally constant states, $i$ = (a, b, c, d),

$$W_{OUT,i} = Q_{ABS,i} = Q_{IN,a} = Q_{IN,b} = W_{IN,c} = W_{IN,d} = Q_{GEN,c} = Q_{GEN,d}, \tag{E.13}$$

which leads to

$$\Delta S_a = \Delta S_b = \Delta S_c = \Delta S_d . \tag{E.14}$$

This entails that equation (II), $dS = dQ/T$, holds true for not “reversible” phenomena, also.

---

[18] The assumption does not change the conclusion, but simplifies the analysis.

### E.3 Cycles – No Losses

Consider two *batch* (time independent)[19], *lossless* cycles of isothermal compression, adiabatic compression, isothermal expansion and adiabatic expansion, i.e. adiabatic-isothermal cycles, of an ideal gas in a closed system, operating between the same, four spatially uniform, temporally constant states, $\text{state}_1$ – $\text{state}_4$, as in figure E.5.

In case a) with heat transfer with zero temperature difference in the isothermal units $U_1$ and $U_3$.

In case b) with heat generation due to work transfer turning a mixer with non-zero time rate in the isothermal unit $U_3$ and with heat transfer with non-zero temperature difference in the isothermal units $U_1$.[20]

These entail that cycle a) is “reversible”, while cycle b) is not “reversible”, according to Clausius [1-6].

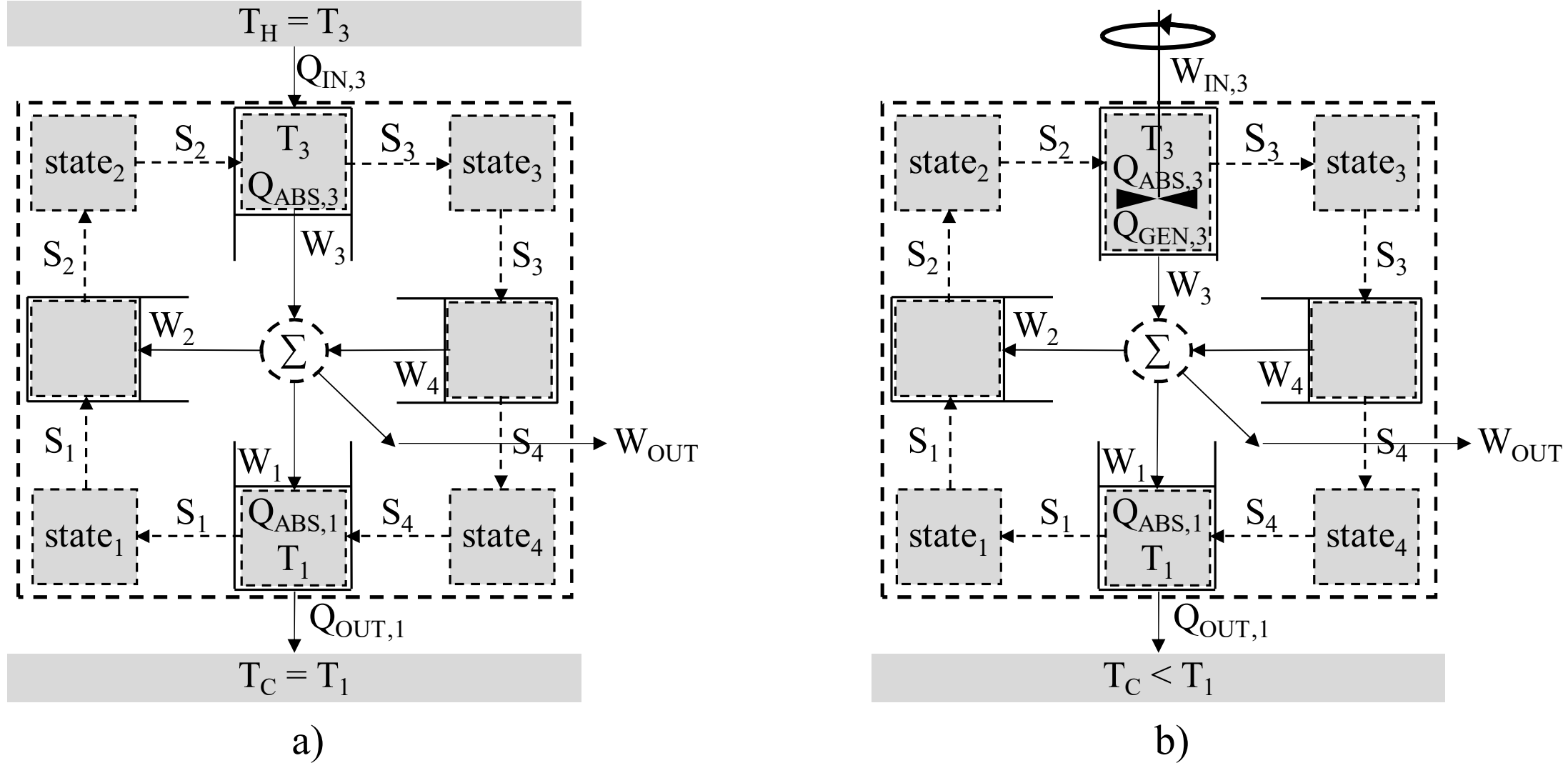

**Figure E.5: Two batch cycles of an ideal gas**

For simplicity, assume uniform temperature and pressure of the working fluid in the units.[21]

Given $Q_{IN,3,a} = W_{IN,3,b}$, $Q_{OUT,1,a} = Q_{OUT,1,b}$, no lossless, and adiabatic-isothermal operation, the four states, $\text{state}_1$ – $\text{state}_4$, of the *working fluid* are exactly the same in cases a) and b) – as was initially specified – independently of “reversibility” and the temperatures of the two heat “reservoirs”.

Thus, if equation (II), $dS = dQ/T$, holds true for the “reversible” units $U_{1,a}$ and $U_{3,a}$, equation (II), $dS = dQ/T$, holds true for the not “reversible” units $U_{1,b}$ and $U_{3,b}$, also.

This entails that equation (II) holds true for not “reversible” phenomena, also.

This entails, further, that equation

$$\oint \frac{dQ}{T} = 0\,, \tag{E.15}$$

holds true for not “reversible” cyclical processes, also – contrary to Clausius [4], p. 129, [6], p. 89.

---

[19] The analysis with steady (time independent) cycles gives the same result. A steady cycle is presented in figure E.6.

[20] In case b), there is heat generation in the isothermal unit $U_3$ – but no losses in the cycle.

[21] The assumption does not change the conclusion, but simplifies the analysis.

## E.4 Cycles – Losses

Consider the *steady* (time independent), *adiabatic-isothermal* (isothermal compression, adiabatic compression, isothermal expansion and adiabatic expansion) cycle in figure E.6. Note that the temperatures of the heat “reservoirs” are not specified..

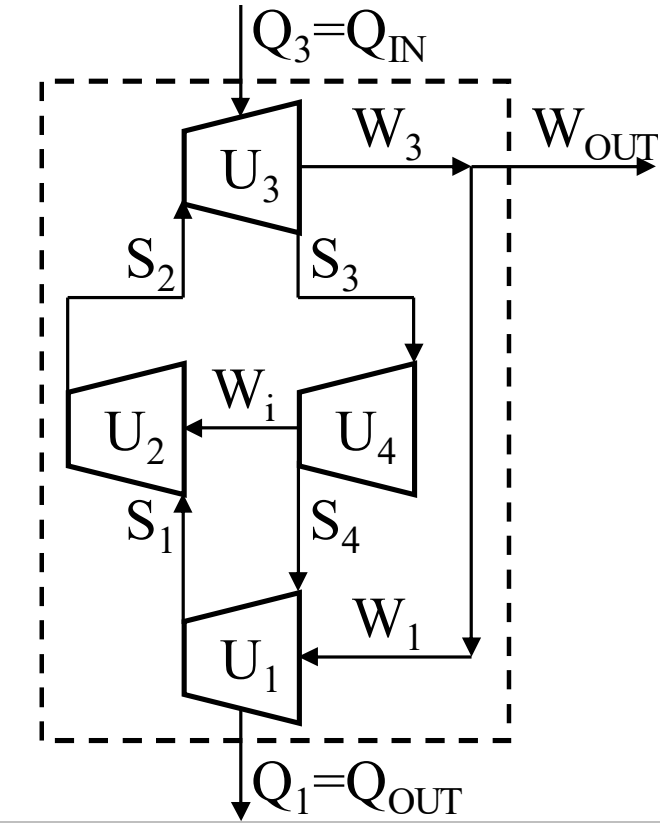

**Figure E.6: A steady adiabatic-isothermal cycle of an ideal gas**

Consider operation with 1 mol/s of an ideal gas with k = $c_p/c_V$ =1.4 with isothermal compression from (1 bara, 25 C, 1 J/K/mol) to 3 bara and adiabatic compression to 9 bara. Assume the pressure of stream $S_3$ after the isothermal expansion adjusted such that $|W_2|=|W_4|$.

Consider operation, first, *ideally*, i.e. without losses, and, second, *non-ideally*, i.e. with losses, in the case of both the isothermal compression in unit $U_1$ and the adiabatic compression in unit $U_2$ having an efficiency of 0.9.

For simplicity, assume radially uniform temperature and pressure of the working fluid.

### *E.4.1 Work, Heat and Temperature Equations*

The work inputs $W_{IN}$ into an *open* system of *ideal* adiabatic and isothermal compressions, respectively, of an ideal gas from $p_1$, $T_1$ to $p_2$, using $k = c_p/c_V$, are

$$\tilde{W}_{IN,\Delta p,adiabatic,ideal} = \frac{\tilde{n}RT_1}{\left(\frac{k-1}{k}\right)}\left[\left(\frac{p_2}{p_1}\right)^{\left(\frac{k-1}{k}\right)} - 1\right], \tag{E.16}$$

$$\tilde{W}_{IN,\Delta p,isothermal,ideal} = \tilde{n}RT_1 \ln\left(\frac{p_2}{p_1}\right). \tag{E.17}$$

The losses are considered by the compression efficiency $\eta_C$ defined as

$$\frac{1}{\eta_C}\tilde{W}_{IN,ideal} \equiv \tilde{W}_{IN} = \left(\tilde{W}_{IN,ideal} + \tilde{W}_{IN,loss}\right) = \left(\tilde{W}_{IN,ideal} + \tilde{Q}_{GEN}\right), \tag{E.18}$$

leading to

$$\tilde{Q}_{GEN} = \tilde{W}_{IN,loss} = (1-\eta_C)\frac{1}{\eta_C}\tilde{W}_{IN,ideal}. \tag{E.19}$$

The heat input $Q_{IN}$ is zero in adiabatic compression and equals the additive inverse of the work input in isothermal compression, i.e.

$$\tilde{Q}_{IN,\Delta p,adiabatic} \equiv 0\,, \tag{E.20}$$

$$\tilde{Q}_{IN,\Delta p,isothermal} = -\tilde{W}_{IN}\,. \tag{E.21}$$

The temperature $T_{OUT}$ of a unit is calculated from the *steady state* energy balance of an *open system*, which, for an ideal gas with constant molar heat capacity, gives

$$T_{OUT} = T_{IN} + \frac{\tilde{Q}_{IN} + \tilde{W}_{IN}}{\tilde{n} C_{p,n}}\,. \tag{E.22}$$

### *E.4.2 Entropy Equations*

In *both ideal and non-ideal* isothermal compression and expansion, the heat absorbed $Q_{ABS}$ equals the additive inverse of the ideal work input, because the heat generated by loss of work is removed from the system (or first absorbed and then desorbed) with no net effect on energy or entropy. Thus, the entropy generation, based on equation (II), $dS = dQ/T$, is

$$\tilde{S}_{GEN,\Delta p,isothermal} = \int_0^{\tilde{Q}} \frac{d\tilde{Q}_{ABS}}{T} = \frac{\tilde{Q}_{ABS}}{T} = \frac{-\tilde{W}_{IN,ideal}}{T}\,. \tag{E.23}$$

In *ideal*, adiabatic compression, the heat generation is zero and thus the entropy change is zero. In *non-ideal*, adiabatic compression, the heat absorbed equals the heat generated, which heat absorption is manifested as a temperature increase over the ideal temperature increase. Thus, the entropy generation, based on equation (II), $dS = dQ/T$, is

$$\tilde{S}_{GEN,\Delta p,adiabatic} = \int_0^{\tilde{Q}} \frac{d\tilde{Q}_{ABS}}{T} = \int_0^{\tilde{Q}} \frac{d\tilde{Q}_{GEN}}{T} = \int_{T_{2,ideal}}^{T_2} \frac{\tilde{n} C_{p,n} dT}{T} = \tilde{n} C_{p,n} \ln\left(\frac{T_2}{T_{2,ideal}}\right). \tag{E.24}$$

### *E.4.3 Results*

The results calculated by the equations (E.16)-(E.24) are presented in tables E.1-E.4 and figure E.7.

| Stream | p | T | S,n |
|---|---|---|---|
| - | bara | C | J/K/mol |
| S4,i | 1.00 | 25 | 1.00 |
| S1 | 3.00 | 25 | - 8.13 |
| S2 | 9.00 | 135 | - 8.13 |
| S3 | 3.00 | 135 | 1.00 |
| S4,f | 1.00 | 25 | 1.00 |

Input specifications grayed.

**Table E.1: Streams, ideal.**

| Unit | eff | W,IN | Q,IN | Q,GEN | S,GEN |
|---|---|---|---|---|---|
| - | - | kW | kW | kW | J/K/s |
| U1 | 1.00 | 2.72 | - 2.72 | 0.00 | - 9.13 |
| U2 | 1.00 | 3.20 | 0.00 | 0.00 | 0.00 |
| U3 | 1.00 | - 3.73 | 3.73 | 0.00 | 9.13 |
| U4 | 1.00 | - 3.20 | 0.00 | 0.00 | 0.00 |
| U,sum | - | - 1.00 | 1.00 | 0.00 | 0.00 |

Input specifications grayed.

**Table E.2: Units, ideal.**

| Stream | p | T | S,n |
|---|---|---|---|
| - | bara | C | J/K/mol |
| S4,i | 1.00 | 25 | 1.00 |
| S1 | 3.00 | 25 | - 8.13 |
| S2 | 9.00 | **147** | **- 7.28** |
| S3 | **3.33** | **147** | 1.00 |
| S4,f | 1.00 | 25 | 1.00 |

Changes relative to ideal case boldfaced

**Table E:3: Streams, non-ideal.**

| Unit | eff | W,IN | Q,IN | Q,GEN | S,GEN |
|---|---|---|---|---|---|
| - | - | kW | kW | kW | J/K/s |
| U1 | **0.90** | **3.03** | **- 3.03** | **0.30** | - 9.13 |
| U2 | **0.90** | **3.55** | 0.00 | **0.36** | **0.86** |
| U3 | 1.00 | **- 3.48** | **3.48** | 0.00 | **8.28** |
| U4 | 1.00 | **- 3.55** | 0.00 | 0.00 | 0.00 |
| U,sum | - | **- 0.45** | **0.45** | **0.66** | 0.00 |

Changes relative to ideal case boldfaced

**Table E.4: Units, non-ideal.**

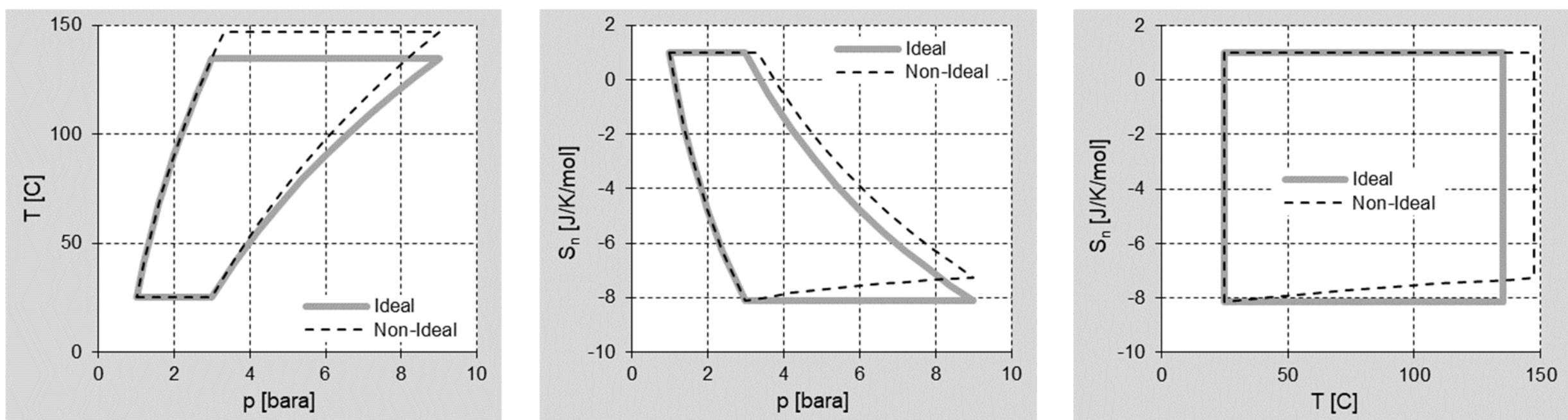

**Figure E.7: T,p-, S,p-, and S,T-cycles**

It is seen that – for both the ideal and non-ideal operations – the initial and final properties of stream S4 are equal, which shows that the operations are truly cyclic.

### *E.4.4 Conclusions*

According to Clausius [4], p. 129, [6], p. 89, equation (E.15)

$$\oint \frac{dQ}{T} = 0\,, \tag{E.15}$$

holds true for “reversible” phenomena, i.e. for *zero temperature difference* and *no heat generation*, *only*.

Now, the use of equations (E.23) and (E.24) – based on equation (II), $dS = dQ/T$ – for the ideal (with no losses) case and the non-ideal (with losses) case of the cycle in figure E.6 gives

$$\begin{aligned} \oint \frac{d\tilde{Q}}{T} = \sum_{i=1}^{4}\left(\frac{\tilde{Q}_{ABS}}{T}\right)_{Ui} &= \frac{-\tilde{W}_{IN,U1,ideal}}{T_{U1}} + 0 + \frac{-\tilde{W}_{IN,U3,ideal}}{T_{U3}} + 0 \\ &= \left[\frac{-2720}{(25+273)} + 0 + \frac{3730}{(135+273)} + 0\right] \text{J/K/s} \\ &= 0\ \text{J/K/s} \end{aligned}\quad, \tag{E.25}$$

$$\begin{aligned} \oint \frac{d\tilde{Q}}{T} = \sum_{i=1}^{4}\left(\frac{\tilde{Q}_{ABS}}{T}\right)_{Ui} &= \frac{-\tilde{W}_{IN,U1,ideal}}{T_{U1}} + \tilde{n}C_{p,n}\int_{T_{S2,ideal}}^{T_{S2}} \frac{dT}{T} + \frac{-\tilde{W}_{IN,U3,ideal}}{T_{U3}} + 0 \\ &= \left[\frac{-2720}{(25+273)} + 1.0\cdot 29.1\cdot \ln\left(\frac{147+273}{135+273}\right) + \frac{3480}{(147+273)} + 0\right] \text{J/K/s}\,. \\ &= 0\ \text{J/K/s} \end{aligned} \tag{E.26}$$

With non-zero temperature difference implied in figure E.6 and heat generation in the non-ideal case, these entail that equation (II) and equation (E.15) hold true for not “reversible” phenomena, also.

# APPENDIX F: ENTROPY MODELING WITHOUT "REVERSIBILITY"

## F.1 Cycle Efficiencies

The efficiency of a heat engine is defined as

$$\eta_{HE} \equiv \frac{W_{OUT}}{Q_{IN}}. \tag{F.1}$$

The energy balance – based on equation (I) – of a *steady, adiabatic-isothermal cycle*, as in figure E.6, is

$$\frac{dE}{dt} = 0 = \tilde{Q}_3 - \tilde{Q}_1 - \tilde{W}_{OUT} = \tilde{Q}_{IN} - \tilde{Q}_{OUT} - \tilde{W}_{OUT}, \tag{F.2}$$

which gives

$$\eta_{HE} \equiv \frac{\tilde{W}_{OUT}}{\tilde{Q}_{IN}} = \frac{\tilde{Q}_{IN} - \tilde{Q}_{OUT}}{\tilde{Q}_{IN}} = 1 - \frac{\tilde{Q}_{OUT}}{\tilde{Q}_{IN}}. \tag{F.3}$$

The entropy balance – based on equation (II) – of a *steady, ideal (lossless)*, *adiabatic-isothermal cycle*, is

$$\frac{dS}{dt} = 0 = \frac{\tilde{Q}_3}{T_3} - \frac{\tilde{Q}_1}{T_1} = \frac{\tilde{Q}_{IN}}{T_{\max}} - \frac{\tilde{Q}_{OUT}}{T_{\min}}, \tag{F.4}$$

which gives, further,

$$\eta_{HE} = 1 - \frac{T_{\min}}{T_{\max}}. \tag{F.5}$$

Any heat absorbed at a temperature of the working lower that $T_3 = T_{\max}$ and any heat desorbed at a temperature of the working higher that $T_1 = T_{\min}$, would decrease the efficiency.

Accordingly, the maximum efficiency of any heat engine is

$$\eta_{HE,\max} = 1 - \frac{T_{\min}}{T_{\max}}. \tag{F.6}$$

Accordingly, the two *necessary and sufficient* conditions of the maximum efficiency of a heat engine are:

1. *Adiabatic-isothermal cycle*: All heat transfer at the extremum temperatures of the working fluid.
2. *Ideality*: No losses of mechanical energy to heat and no heat losses in or out of the system.

As a numeric example, for the cycle in figure E.6 the values in Tables E.1-E.4 give:

| | *ideal* | *non-ideal* | |
|---|---|---|---|
| $W_{\mathrm{OUT}}$ | 1.00 | 0.45 | kW |
| $Q_{\mathrm{IN},3}$ | 3.73 | 3.48 | kW |
| $\eta_{\mathrm{HE}} = W_{\mathrm{OUT}}/Q_{\mathrm{IN},3}$ | 0.27 | 0.13 | - |
| $T_{\max}$ | 135 | 147 | C |
| $T_{\min}$ | 25 | 25 | C |
| $\eta_{\mathrm{HE,max}} = 1 - T_{\min}/T_{\max}$ | 0.27 | 0.29 | - |
| $\eta_{\mathrm{HE}}/\eta_{\mathrm{HE,max}}$ | 1.00 | 0.45 | - |

**Table F.1: Efficiencies for the Cycle in Figure E.6**

## F.2 Inequalities and the Principle of the Increase of Entropy

Because “reversibility” is false, the inequalities, e.g. Clausius [4], p. 329, [6], p. 213,

$$\oint \frac{dQ}{T} \leq 0 \,, \tag{F.7}$$

are false and the corresponding equalities are true for all phenomena, whether “reversible” or not.

However, consider an *isolated system* with two closed bodies A and B with uniform, but not constant temperatures, $T_A$ and $T_B$, which gives,

$$\frac{dS}{dt} = \tilde{S}_{GEN} = \frac{\tilde{Q}_{ABS,A}}{T_A} + \frac{\tilde{Q}_{ABS,B}}{T_B} \geq 0 \,, \tag{F.8}$$

in which the leftmost equality follows from the isolated system and the middle equality follows from the use of equation (II), $dS = dQ/T$ for bodies A and B, separately.

Because of isolated system and energy conservation,

$$\left|\tilde{Q}_{ABS,A}\right| = \left|\tilde{Q}_{ABS,B}\right| . \tag{F.9}$$

Now, in equation (F.8), the rightmost equality follows from $T_A = T_B$ leading to $\tilde{Q}_{ABS,A} = \tilde{Q}_{ABS,B} = 0$ and the inequality follows from $T_A > T_B$ leading to $\tilde{Q}_{ABS,A} < 0 \,, \tilde{Q}_{ABS,B} > 0$ .

The result may be *assumed* to hold true for any phenomena, *in general*, as [22]

$$\left(\frac{dS}{dt}\right)_{isolated} = \left(\tilde{S}_{GEN}\right)_{isolated} \geq 0 \,, \tag{F.10}$$

i.e.,

*Entropy accumulation and entropy generation are non-negative in an isolated system*. (F.11)

---

[22] This generalization is generalized, further, by Clausius [4], p. 356, to the latter of:
“ 1. *The energy of the universe is constant.*
2. *The entropy of the universe tends to a maximum*. ”

## F.3 Equilibrium

### *F.3.1 "Reversibility"*

According to Clausius [1-6], "reversible" are only phenomena of *zero temperature difference* – i.e. of *temperature equilibrium*.

This *may explain* the use of the attributes "reversible" and "equilibrium" synonymously in the context of entropy modeling and the conception that "equilibrium" is a necessary condition for "reversibility" and for the equation (II), $dS = dQ/T$, to hold true.

However, because the domain of the equation (II), $dS = dQ/T$, is not restricted to "reversible" phenomena, the domain of the equation (II) is not restricted to "equilibrium" phenomena.

Within the Onsager-Prigogine entropy model, the issue is dealt with assuming "local equilibrium", e.g., de Groot and Mazur [9], p. 23 [underline added]:

> "It will now be assumed that, although the total system is not in equilibrium, there exists within small mass elements a state of 'local' equilibrium, for which the local entropy $s$ is the same function […] of $u$, $v$ and $c_k$ as in real equilibrium."
> "This hypothesis of 'local' equilibrium can, from a macroscopic point of view, only be justified by virtue of the validity of the conclusions derived from it."

The adequacy of the assumption of "local equilibrium" in the modeling of *any given phenomena* must depend on the *physical reality* and not on the conceptual modeling context.

### *F.3.2 Reality*

According to Landau and Lifshitz [11], p. 6, statistical/thermodynamic/thermal equilibrium refers to a state of a macroscopic system such that "in any macroscopic subsystem the macroscopic physical quantities are to a high degree of accuracy equal to their mean values".

Now, this may be seen to entail a fundamental problem, e.g., Moran et al. [12], p. 190 [underline added]:

> "Although each end state is an equilibrium state at the same pressure and temperature, the pressure and temperature are not necessarily uniform throughout the system at intervening states, nor are they necessarily constant in value during the process. Accordingly, there is no well-defined 'path' for the process."

However, (*continuous, differential) field equations* for "[m]otion, stress, energy, entropy, and electromagnetism" may be justified as "phenomenological", Truesdell and Toupin [13], pp. 226-233. With or without explicit justification, field equations are ubiquitously used and found adequate in the context of continuum physics.[23]

Obviously, continuum physics has limitations in modeling the non-continuous physical reality, but this topic, in general, is out of the scope, here.

---

[23] See, e.g., equations (E.5) and (E.6) in example E.1 Heating in appendix E.